\documentclass[prl,amssymb,amsmath,amsfonts,floatfix,onecolumn]{revtex4}
\usepackage{times}
\usepackage{color}
\begin{document}



\title[Conformal Transformations in Metric-Affine Gravity]{Conformal Transformations in
Metric-Affine Gravity and Ghosts}


\author{Canan N. Karahan$^{a}$, Oktay Do{\u g}ang{\"u}n$^{b}$, Durmu{\c s} A. Demir$^{a}$ }

\address{$^{a}$ Department of Physics, Izmir Institute of Technology, Izmir, TR35430, Turkey \\
$^{b}$ Department of Physical Sciences, University of Naples \& INFN, Naples, Italy}

\begin{abstract}
Conformal transformations play a widespread role in gravity theories in regard
to their cosmological and other implications. In the pure metric theory of gravity,
conformal transformations change the frame to a
new one wherein one obtains a conformal-invariant scalar-tensor theory such that
the scalar field, deriving from the conformal factor, is a ghost.

In this work, conformal transformations and ghosts will be analyzed in the framework
of the metric-affine theory of gravity. Within this framework,
metric and connection are independent variables, and hence, transform independently
under conformal transformations. It will be shown that, if affine connection is invariant
under conformal transformations then the scalar field under concern is a non-ghost, non-dynamical field.
It is an auxiliary field at the classical level, and might develop a kinetic term at the quantum level.

Alternatively, if connection transforms additively with a structure similar to
yet more general than that of the Levi-Civita connection, the resulting
action describes the gravitational dynamics correctly, and  more importantly,
the scalar field becomes a dynamical non-ghost field. The equations of
motion, for generic geometrical and matter-sector variables, do not reduce
connection to the Levi-Civita connection, and hence, independence of connection from
metric is maintained. Therefore, metric-affine gravity provides an arena in which
ghosts arising from conformal factor are avoided thanks to the independence
of connection from the metric.
\end{abstract}

\maketitle

\section{Introduction}
Spacetime is a smooth manifold ${\cal{M}}\left(g,\Gamma\right)$ endowed with a
metric $g$ and connection $\Gamma$. Metric is responsible for measuring the
distances while connection governs curving and twirling of the manifold. Connection
specifies how vectors and tensors are to be differentiated in curved spacetime.
We hereby emphasize that only the symmetric connections {\it i.e.} torsion-free spacetimes will be considered throughout
in this work. These two geometrical structures, the metric and connection, are fundamentally independent
geometrical variables, and they play completely different roles in spacetime dynamics. If they are to exhibit any relationship
it derives from dynamical equations {\it a posteriori}. This fact, that the
metric and connection are independent geometrical variables, gives rise to two
alternative approaches to General Relativity (GR):
\begin{enumerate}
\item GR with metricity only\,,
\item GR with affinity and metricity\,.
\end{enumerate}
The former is a purely metric theory of gravity since connection is completely
determined by the metric and its partial derivatives, {\it a priori}. This
determination is encoded in the Levi-Civita connection, $\check{\Gamma}$ as
\begin{eqnarray}
\label{levi-civita}
\check{\Gamma}^{\lambda}_{\alpha\beta} = \frac{1}{2} g^{\lambda \rho}
\left(\partial_{\alpha} g_{\beta \rho} + \partial_{\beta} g_{\rho\alpha} -
\partial_{\rho} g_{\alpha\beta}\right)\,,
\end{eqnarray}
which defines a metric-compatible covariant derivative \cite{book1}. In this
particular setup, the Einstein-Hilbert action produces gravitational field
equations only by adding an extrinsic curvature term.

The metric-affine gravity (similar to the Palatini formalism\cite{palatini,palatini-pheno,higher-palatini}
in philosophy), which considers an independence of metric tensor and connection
\cite{book1,reviews}, encodes a more general approach to gravitation by breaking
up the {\it a priori} relation (\ref{levi-civita}). In this approach,
gravitational field equations are successfully produced with no need to
extrinsic curvature term provided that the geometrical sector is minimal and
matter sector is independent of the connection \cite{reviews}. Concerning
the matter sector, the fermion kinetic term \cite{deser1} and coupling between
scalar fields and curvature scalar \cite{bauer1} are just two examples that
immediately come to mind.

This work is about yet another difference between the metrical and affine
approaches to gravity. Conformal transformation is essentially a local change of
scale. Since distances are measured by metric, such transformations executed by
rescaling the metric by a smooth, nonvanishing and space-time dependent function
$\Omega(x)$, called the conformal factor \cite{weyl,book1}. Therefore, the
transformation
\begin{eqnarray}
\label{transmet}
\widetilde{g}_{\alpha \beta} = \Omega^2(x)\ g_{\alpha \beta}
\end{eqnarray}
which shrinks or stretches the distances on the manifold locally.

Conformal transformations are particularly respectful to
distinction between metric and connection. Indeed, transformation
of the metric in (\ref{transmet}) automatically induces a transformation
of the Levi-Civita connection as
\begin{eqnarray}
\label{trans-conn}
\widetilde{\check{\Gamma}}^{\lambda}_{\alpha \beta} =
\check{\Gamma}^{\lambda}_{\alpha \beta} +
\Delta^{\lambda}_{\alpha\beta}
\end{eqnarray}
with
\begin{eqnarray}
\label{Delta-for-Levi-Civita}
\Delta^{\lambda}_{\alpha\beta} = \delta^{\lambda}_{\beta}
\partial_{\alpha}\ln\Omega
+ \delta^{\lambda}_{\alpha} \partial_{\beta}\ln\Omega - g_{\alpha\beta}
\partial^{\lambda}\ln\Omega\,.
\end{eqnarray}
However, this direct correlation is completely lost in the metric-affine gravity
since there is no telling of
how the general connection
\begin{eqnarray}
\label{pure connection}
\Gamma^{\lambda}_{\alpha \beta} \neq \check{\Gamma}^{\lambda}_{\alpha\beta}(g)
\end{eqnarray}
should transform under a rescaling of distances. In fact, the fact that
connection has nothing to
do with measuring the distances can be taken to imply that the connection
$\Gamma^{\lambda}_{\alpha\beta}$ is completely inert under (\ref{transmet}).
However, it
is still possible that connection still transforms in some way, not necessarily
like (\ref{trans-conn}).
Stating in a clearer fashion, there arise two main categories to be explored:
\begin{itemize}
\item The connection $\Gamma$ can be conformal-invariant:
$\Gamma^{\lambda}_{\alpha\beta} \rightarrow \Gamma^{\lambda}_{\alpha\beta}$
despite (\ref{transmet}) \cite{weyl},
\item The connection $\Gamma$ can transform in various ways: Multiplicatively,
additively or both while metric transforms as in (\ref{transmet}).
\end{itemize}
Each of these two possibilities gives rise to novel effects not found in
metrical GR,
as indicated by the dependence of the Riemann curvature tensor on the connection
\begin{eqnarray}
\label{riemann}
\mathbb{R}^{\alpha}_{\mu \beta \nu}\left(
\Gamma\right) =
\partial_{\beta} \Gamma^{\alpha}_{\mu \nu} - \partial_{\nu}
\Gamma^{\alpha}_{\mu \beta} + \Gamma^{\alpha}_{\beta \lambda}
\Gamma^{\lambda}_{\mu \nu} - \Gamma^{\alpha}_{\nu \lambda}
\Gamma^{\lambda}_{\mu \beta}\,.
\end{eqnarray}
It is obvious that the two conformal transformation categories mentioned above
will, in general, lead to completely new dynamics with no analogue in metrical
GR. This work is devoted to a comparative analysis of conformal transformations
in the GR with metricity and GR with affinity in the pathways described by these
two categories.

In the body of the work below, we first give a discussion of metrical GR.
Following this
we turn to a detailed analysis of the metric-affine gravity by exploring
plausible alternatives one by
one. After this, in the last section, we discuss certain salient features of the
model
not covered in the text and conclude.
\section{Conformal Transformations in GR with Metricity}
\label{sec:GR}
In metrical GR, conformal transformation of the metric (\ref{transmet})
automatically
leads to transformation of the connection (\ref{trans-conn}), and hence, of the
Riemann curvature tensor (\ref{riemann}). The transformed Riemann tensor reads
as
\begin{eqnarray}
\widetilde{R}^\alpha_{\mu\beta\nu}\left(\widetilde{\check{\Gamma}}\right) &=&
R^\alpha_{\mu\beta\nu}\left(\check{\Gamma}\right)
+ \left[ \delta_\beta^\alpha \delta_\nu^\lambda \delta_\mu^\rho -
\delta_\nu^\alpha \delta_\beta^\lambda \delta_\mu^\rho
 +  \delta_\beta^\lambda g_{\mu \nu} g^{\alpha \rho} \right. \nonumber \\
\label{transRiemanntensor}
&\quad & - \left. \delta_\nu^\lambda g_{\mu \beta} g^{\alpha \rho} \right]
\left( 2 \partial_\lambda \ln \Omega\ \partial_\rho \ln \Omega - \nabla_\lambda
\partial_\rho \ln \Omega \right) \nonumber\\
&\quad& + \left[  \delta_\nu^\alpha g_{\mu
\beta} g^{\lambda \rho} -  \delta_\beta^\alpha g_{\mu \nu} g^{\lambda \rho}
\right] \partial_\lambda \ln \Omega\ \partial_\rho \ln \Omega
\end{eqnarray}
where use has been made of the definition
$\mathbb{R}^{\alpha}_{\mu\beta\nu}\left(\Gamma = \check{\Gamma}\right) \equiv
R^{\alpha}_{\mu\beta\nu}\left(\check{\Gamma}\right)$.
Contraction of this rank (1,3) tensor gives the transformed Ricci tensor
\begin{eqnarray}
\label{transRiccitensor}
\widetilde{R}_{\mu\nu}\left(\widetilde{\check{\Gamma}}\right) &=&
R_{\mu\nu}\left(\check{\Gamma}\right) - \left[ (D-2) \nabla_{\mu}\nabla_{\nu}
+ g_{\mu\nu} \Box\right] \ln \Omega \nonumber \\ &+& \left[ 2 (D-2)
\delta_\mu^\alpha
\delta_\nu^\beta - (D-3) g_{\mu \nu} g^{\alpha \beta} \right] \partial_\alpha
\ln \Omega\ \partial_\beta \ln \Omega
\end{eqnarray}
so that transformed Ricci scalar takes the form
\begin{eqnarray}
\widetilde{R}\left(\widetilde{\check{\Gamma}}\right) &=& \widetilde{g}^{\mu\nu}
\widetilde{R}_{\mu \nu}\left(\widetilde{\check{\Gamma}}\right) \nonumber\\ &=&
\Omega^{-2} \left[R\left(\check{\Gamma}\right) - 2 (D-1) \Box \ln \Omega
- (D-1) (D-2)
g^{\alpha\beta} \partial_\alpha
\ln \Omega \partial_\beta \ln \Omega \right]\,. \nonumber\\
\end{eqnarray}
These transformation properties dictate how gravitational action density
transforms under conformal
rescalings. The Einstein-Hilbert action reads in $\left(g,\check{\Gamma}\right)$
frame
as
\begin{eqnarray}
\label{eq:grav-ac}
S_{EH}\left[g\right] &= \int d^{D}x\ \sqrt{-g} \left[\frac{1}{2} M_{\star}^{D-2}
R - \Lambda_{\star} + \mathfrak{L}_{m}\left(g,\Psi\right) \right]
\end{eqnarray}
where $M_{\star}$ is the fundamental scale of gravity in $D$
dimensions, $\Lambda_{\star}$ is the cosmological term, and $\mathfrak{L}_{m}$
is the Lagrangian of the matter and radiation fields, collectively denoted by $\Psi$. For the metric
$(-,+,\dots,+)$ convention is adopted. Under the conformal transformation of the
metric (\ref{transmet}), this action becomes in $\left(\widetilde{g},
\widetilde{\check{\Gamma}}\right)$ frame
\begin{eqnarray}
\label{eq:GR-action}
S_{EH} \left[g, \overline{\phi}\right] &=& \int d^{D}x\ \sqrt{-g}
\Bigg\{\frac{1}{2} M_{\star}^{D-2} \left[ \Omega^{D-2} R - 2 (D-1) \Omega^{D-2}
\Box \ln \Omega \right. \nonumber \\
&\quad& - \left. (D-1) (D-2) \Omega^{D-4} g^{\alpha\beta} \nabla_{\alpha}
\Omega
\nabla_{\beta} \Omega \right] - \Lambda_{\star} \Omega^D  + \widetilde{\mathfrak{L}}_{m}\left(g,\widetilde{\Psi}\right) \Bigg\}\nonumber\\
&\equiv&  \int d^{D} x\ \sqrt{{g}} \left[
\frac{1}{2} g^{\mu \nu} \left({\partial}_{\mu} \overline{\phi}\right)
\left(\partial_{\nu} \overline{\phi}\right) + \frac{1}{2} \zeta_D
\overline{\phi}^2 {R} - \lambda_D \left(\zeta_D
\overline{\phi}^2 \right)^{\frac{D}{D-2}} + \widetilde{\mathfrak{L}}_{m}\left(g,\widetilde{\Psi}\right) \right] \nonumber \\
\label{transac}
\end{eqnarray}
where the two dimensionless constants
\begin{eqnarray}
\label{eq:newquantities}
\zeta_D = \frac{D-2}{4 (D-1)}\,,\; \lambda_D =
\frac{\Lambda_{\star}}{M_{\star}^D}
\end{eqnarray}
designate, respectively, the conformal coupling of $\overline{\phi}$ to $R$
and the self-coupling of $\overline{\phi}$. The scalar field $\overline{\phi}$
\begin{eqnarray}
\overline{\phi} = \frac{1}{\sqrt{\zeta_D}}\ \left({M}_{\star}
{\Omega}\right)^{\frac{(D-2)}{2}}
\end{eqnarray}
derives from the conformal factor $\Omega$  in order to
have canonical kinetic term. The quantity $\widetilde{\mathfrak{L}}_{m}\left(g,\widetilde{\Psi}\right)$ in (\ref{eq:GR-action})
is the transformed matter Lagrangian, where each matter field $\Psi$ transforms, together with the metric,
by an appropriate conformal weight. The conformal weights of fields (charges of fields under scalings) are
determined from (global) conformal invariance of their kinetic terms \cite{weight1,weight2}.

There are certain salient features of the transformed action
(\ref{eq:GR-action}),
which deserve a detailed discussion.
\begin{itemize}
\item First of all, this action executes local conformal invariance (the Weyl
invariance) under
the transformations
\begin{eqnarray}
\label{transmetphi}
{g}_{\alpha \beta} \longrightarrow \psi^{2}\ {g}_{\alpha \beta}\;,\;\;
\overline{\phi} \longrightarrow \psi^{-\frac{(D-2)}{2}}\ \overline{\phi}
\end{eqnarray}
where inhomogeneous terms generated by the kinetic term of $\overline{\phi}$ are
neutralized by the terms generated by the transformation of the curvature scalar
${R}$.
This happens thanks to the special, conformal value of $\zeta_D$. Therefore,
the transformed action (\ref{transac}) provides a locally conformal-invariant
representation of the original Einstein-Hilbert action (\ref{eq:grav-ac}).
Notably, the original
action (\ref{eq:grav-ac}) exhibits no sign of conformal invariance but the
transformed one does
and the reason behind it is the
dressing of
$M_{\star}$ and $\Lambda_{\star}$ by the transformation field $\Omega$
\cite{bekenstein,deser2}.
\item
One can  also notice that; conformal transformation, like Gauge transformations, adds a new degree of freedom to the system. This is a built-in property of the system; this is common to 'transformations' including the gauge transformations.
\item Another point to notice about (\ref{eq:GR-action}) is that the scalar
field $\overline{\phi}$
(which is a function of the conformal factor $\Omega$) is a ghost
\cite{demir,demir2,hawking}. This is an unavoidable feature
if gravity is to be an attractive force. Its ghosty nature follows from its
non-positive kinetic
term, and it signals that the system has no lower bound for energy. Such systems
are inherently
unphysical, and there seems to be no way of avoiding it unless some
nonlinearities are added as
extra features \cite{gabadadze,mazumdar1,mazumdar2}.

\item The transformed action (\ref{eq:GR-action}), when
$\widetilde{g}_{\alpha\beta}=\eta_{\alpha\beta}$,
leads to a $\overline{\phi}^{4}$ theory in $D=4$ dimensions. In this particular
case spacetime is flat, and
entire gravitational effects reduce to a conformal-invariant scalar field
theory. This theory has been
argued to exhibit an infrared fixed point at $\lambda_D = 0$, and this feature
has been suggested to provide
a solution to the cosmological constant problem ($\lambda_D$ is proportional to
the vacuum energy
density $\Lambda_{\star}$ in $D$ dimensions) \cite{polyakov}.
\end{itemize}
\section{Conformal Transformations in GR with Affinity}
\label{sec:MAG}
As mentioned in Introduction, GR with affinity treats metric
and connection as independent geometric variables, as they indeed are. One of
the most
important consequences of this feature is that, conformal transformation of
metric
tensor gives rise to no direct change in connection, as happens in GR with
metricity.
Therefore, parallel to the classification made in Introduction, in this section
we shall
analyze conformal transformations in two separate cases in regard to the
transformation
properties of the connection. In course of the analysis, the main goal will be
to
find appropriate transformation rules for $\Gamma^{\lambda}_{\alpha\beta}$ so
that
the resulting scalar field theory (in terms of the conformal factor $\Omega$)
assumes
physically sensible properties like emergent conformal invariance and absence of
ghosts.
Indeed, the main problem with the metrical GR discussed above is the unavoidable
presence
of a ghosty scalar in the spectrum. We will find that affine GR is capable of
realizing
conformal invariance and accommodating non-ghost scalar degrees of freedom.

In the metric-affine gravity, the Einstein-Hilbert action can be written as
\begin{eqnarray}
S_{EH}\left[g,\Gamma\right] &=& \int d^{D} x\ \sqrt{-g} \Bigg\{\frac{1}{2}
M_{\star}^{D-2} g^{\mu \nu} \mathbb{R}_{\mu \nu}\left(\Gamma\right)
- \Lambda_{\star} + \mathcal{L}\left(\Gamma-\check{\Gamma},
g,\Psi\right) \Bigg\} \nonumber \\
\label{affine-inv-ac}
\end{eqnarray}
in a general $\left(g, \Gamma\right)$ frame. In here, $\Psi$
collectively denotes the matter fields, and ${\cal{L}}$ is composed of
\begin{eqnarray}
\mathcal{L} = \mathcal{L}_{\textrm{geo} }
\left( {g}, \mathcal{D}
\right)
+ \mathfrak{L}_{m} \left( g, \mathcal{D},
\Psi\right)
\end{eqnarray}
which, respectively, stand for the geometrical and matter sector contributions.
The geometrical sector consists of the rank (1,2) tensor field
\begin{eqnarray}
\mathcal{D}_{\alpha\beta}^\lambda = \Gamma_{\alpha\beta}^\lambda -
\check{\Gamma}_{\alpha\beta}^\lambda
\end{eqnarray}
as an additional geometrodynamical tensorial quantity. This variable is
highly natural to consider since in the presence of the metric $g_{\alpha
\beta}$ one naturally defines its compatible connection {\it i. e.} the Levi-Civita
connection. Then
difference between $\Gamma^{\lambda}_{\alpha \beta}$ and Levi-Civita
connection becomes a tensorial quantity to be taken into account.

Here it is useful to clarify the meaning of $\mathcal{L}_{\textrm{geo} }
\left( {g}, \mathcal{D} \right)$ in terms of the known dynamical quantities
akin to non-Riemannian geometries. Non-Riemannian geometries are characterized
by torsion tensor $\mathbb{S}^{\lambda}_{\alpha\beta} = \mathcal{D}_{\alpha\beta}^{\lambda} -
\mathcal{D}_{\beta\alpha}^{\lambda}$, non-metricity $\mathbb{Q}^{\alpha\beta}_{\lambda} = \mathcal{D}^{\alpha}_{\lambda\rho}
g^{\rho \beta} + \mathcal{D}^{\beta}_{\lambda \rho} g^{\alpha \rho}$, the Ricci curvature
tensor $\mathbb{R}_{\mu \nu}\left(
\Gamma\right) =  \mathbb{R}^{\alpha}_{\mu \alpha \nu}\left(
\Gamma\right)$, and the other Ricci curvature tensor $ \mathbb{R}^{\prime}_{\beta \nu} = \mathbb{R}^{\alpha}_{\alpha \beta \nu}\left(
\Gamma\right)$. All these tensor fields make up the geometrical sector of the theory which is obviously span a much larger set compared
to the purely metric formulation (GR). In GR connection and metric are put in direct relation from the scratch. However, physically,
it is more natural to induce a relation between them, if any, as a result of dynamical equations \cite{foundation}. This is what is done by the Palatini formalism where metricity appears in the system automatically via the equations of motion. In metric-affine gravity we explore here metric and connection continue to be independent geometrical variables with no harm from their equations of motion (See Appendix A for further details.). One crucial aspect of non-Riemannian geometries (with non-vanishing torsion and/or nonmetricity) is to provide a compact structuring of various tensor
fields which can play important roles in cosmology \cite{yeni-is}.

Clearly, torsion vanishes for theories with symmetric connections, and this is also the case throughout the present work. Although the torsion-free cases are studied for simplicity, the nonmetricity, which relaxes the restrictions on the theory, still holds. Moreover, $\mathbb{R}^{\prime}_{\beta \nu}$ is an anti-symmetric tensor field whose curvature scalar vanishes identically. This tensor can give contributions to Lagrangian at the quadratic and higher levels. The Lagrangian $\mathcal{L}_{\textrm{geo} }\left( {g}, \mathcal{D} \right)$ includes all these tensorial contributions through the $\mathcal{D}_{\alpha\beta}^\lambda$ dependence
\begin{eqnarray}
\label{express}
\mathcal{L}_{\textrm{geo} }
\left( {g}, \mathcal{D} \right) = \mathcal{L}_{\textrm{geo} }
\left( {g}, \mathbb{S}, \mathbb{Q}, \mathbb{R}, \mathbb{R}^{\prime}\right)\,,
\end{eqnarray}
throughout the text. It is clear that, $\mathcal{L}_{\textrm{geo}}$ can
involve arbitrary powers and derivatives of the tensorial connection
$\mathcal{D}_{\alpha\beta}^\lambda$.

It is clear that the Lagrangian ${\cal{L}}$, through its $\Gamma$ or $\mathcal{D}$ dependence,
gives rise to important modifications in the equations of motion \cite{palatini-variation} so that
$\Gamma=\check{\Gamma}$ limit (which is precisely what is behind the Palatini formulation) does not necessarily
hold. The contributions of $\mathcal{L}_{\textrm{geo} }
\left( {g}, \mathcal{D} \right)$ and $\mathfrak{L}_{m} \left( g, \mathcal{D}, \Psi\right)$ generically
avoid the limit $\Gamma=\check{\Gamma}$. We will discuss this point below. In the following section, however, we
will focus on the transformation properties of  (\ref{affine-inv-ac}) without considering the contributions
of $\mathcal{L}_{\textrm{geo} }
\left( {g}, \mathcal{D} \right)$ or $\mathfrak{L}_{m} \left( g, \mathcal{D}, \Psi\right)$. This is
done for the purpose of definiteness and simplicity.  Nevertheless, in Appendix A, we shall come
back to the effects of $\mathcal{L} $, especially the $\mathcal{L}_{\textrm{geo} }
\left( {g}, \mathcal{D} \right)$, and give a detailed discussion of the equations of
motion and other features.

\subsection{Conformal-Invariant Connection}
\label{sec:weyl-invariant}
We start the analysis by first considering a conformal-invariant connection by
which we mean
that connection is inert to rescalings of the metric. Therefore, along with the
transformation of
metric (\ref{transmet}), the connection transforms as \cite{weyl}
\begin{eqnarray}
\widetilde{\Gamma}^{\lambda}_{\alpha\beta} =
\Gamma^{\lambda}_{\alpha\beta}\,
\end{eqnarray}
and hence,
\begin{eqnarray}
\widetilde{\mathbb{R}}^{\alpha}_{\mu\beta\nu}\left(\widetilde{\Gamma}\right) =
 \mathbb{R}^{\alpha}_{\mu\beta\nu}\left(\Gamma\right)\,,\;
\widetilde{\mathbb{R}}_{\mu\nu}\left(\widetilde{\Gamma}\right) =
\mathbb{R}_{\mu\nu}\left(\Gamma\right)
\end{eqnarray}
since Riemann tensor (\ref{riemann}) does not involve the metric tensor unless
the connection does.
The only non-trivial transformation occurs for the Ricci scalar
\begin{eqnarray}
\label{trans-ricci-scal}
\widetilde{g}^{\mu \nu} \widetilde{\mathbb{R}}_{\mu \nu}
\left(\widetilde{\Gamma}\right) = \Omega^{-2} g^{\mu \nu} \mathbb{R}_{\mu
\nu}\left(\Gamma\right)
\end{eqnarray}
which is nothing but an overall dressing by $\Omega^{-2}$. In particular, no
derivatives of $\Omega$ are
involved in the transformations of curvature tensors. This implies that $\Omega$
can develop no kinetic term. Indeed,
under the transformation (\ref{trans-ricci-scal}), the action
(\ref{affine-inv-ac}) with conformal-invariant connection goes
over
\begin{eqnarray}
\label{affine-inv-ac-trans}
S_{EH}\left[\widetilde{g},\widetilde{\Gamma},\overline{\phi}\right] = \int
d^{D} x\ \sqrt{-g} \left[ \frac{1}{2} \zeta_D \overline{\phi}^2 g^{\mu \nu}
\mathbb{R}_{\mu \nu}\left(\Gamma\right) - \lambda_D \left(\zeta_D
\overline{\phi}^2 \right)^{\frac{D}{D-2}} \right]
\end{eqnarray}
in $\left(\widetilde{g},\widetilde{\Gamma}\right)$ frame  and in
the absence of the geometrical and matter parts $\mathcal{L}$. Obviously, this
action is locally conformal invariant under
\begin{eqnarray}
\label{transmetphix}
{g}_{\alpha \beta} \longrightarrow \psi^{2}\ {g}_{\alpha \beta}\;,\;\;
\Gamma^{\lambda}_{\alpha\beta} \longrightarrow
\Gamma^{\lambda}_{\alpha\beta}\;,\;\;
\overline{\phi} \longrightarrow \psi^{-\frac{(D-2)}{2}}\ \overline{\phi}
\end{eqnarray}
as was the case for metrical gravity, defined in (\ref{transmetphi}). Therefore,
though the original
action (\ref{affine-inv-ac}) exhibits no sign of conformal invariance
and hence the new action (\ref{affine-inv-ac-trans}) arises, this
transformed action exhibits manifest conformal invariance. The reason
is as in the metrical gravity; the conformal factor $\Omega$ dresses
the fixed scales ($M_{\star}$ and $\Lambda_{\star}$) in (\ref{affine-inv-ac})
to make them as effective fields transforming nontrivially under local
rescalings of the fields \cite{bekenstein}.

Apart from this emergent conformal invariance, the action
(\ref{affine-inv-ac-trans})
possesses a highly important aspect not found in metrical GR: It is that
$\overline{\phi}$ is not a ghost at all. It is a non-dynamical scalar field
having vanishing kinetic energy, and thus, the impasse caused by the ghosty
scalar field encountered in metrical GR is resolved. The non-dynamical nature
of $\overline{\phi}$ continues to hold even if the matter sector is included.
This result stems form the affine nature of the gravitational theory under
concern,
and especially from the invariance of the connection under conformal
transformations.

At this point it proves useful to discuss the `non-dynamical' nature of the
scalar field $\overline{\phi}$ in the action (\ref{affine-inv-ac-trans}). At
the level of the transformations employed and the Einstein-Hilbert action
the non-dynamical nature of the conformal factor (and hence, the $\overline{\phi}$)
is unavoidable. However, one immediately notices that this `non-dynamical' structure
depends sensitively on the quantum fluctuations. Indeed, if quantum fluctuations
are included into (\ref{affine-inv-ac-trans}) the scalar field $\overline{\phi}$ is
found to develop a kinetic term via the graviton loops \cite{Shapiro2}. We shall
keep analysis at the classical level throughout the work. However, one is warned of
such delicate effects which can come from quantum corrections or higher order
geometrical invariants.

\subsection{Conformal-Variant Connection}
As an alternative to conformal-invariant connection, in this subsection we
investigate
different scenarios where $\Gamma^{\lambda}_{\alpha \beta}$ exhibits
nontrivial
changes along with the transformation of the metric in (\ref{transmet}).

As a possible transformation property, we first discuss the multiplicative
transformation
of connection. Namely, connection transforms similar to the metric itself
\begin{eqnarray}
\label{trans-metric-multi}
\widetilde{\Gamma}^\lambda_{\alpha\beta} =
f(\Omega)\Gamma^{\lambda}_{\alpha\beta}
\end{eqnarray}
where $f(\Omega)$ is a generic function of the conformal factor. Inserting this
transformed
connection into (\ref{riemann}), one straightforwardly determines the
transformed Riemann tensor
\begin{eqnarray}
\widetilde{\mathbb{R}}^{\alpha}_{\mu\beta\nu}\left(\widetilde{\Gamma}\right)
&=& f\left(\Omega\right)
{\mathbb{R}}^{\alpha}_{\mu\beta\nu}\left({\Gamma}\right) +
\partial_{\beta}f\left(\Omega\right)\Gamma^{\alpha}_{\mu \nu}
 -
\partial_{\nu}f\left(\Omega\right)\Gamma^{\alpha}_{\mu\beta} \nonumber\\
&+& f\left(\Omega\right)\left(f\left(\Omega\right)-1\right)
\left[\Gamma_{\beta \lambda}^{\alpha}\Gamma_{\mu \nu}^{\lambda} - \Gamma_{\nu
\lambda}^{\alpha}\Gamma_{\mu \beta}^{\lambda}\right]
\end{eqnarray}
and hence the transformed Ricci scalar
\begin{eqnarray}
\widetilde{g}^{\mu\nu} \widetilde{\mathbb{R}}_{\mu
\nu}\left(\widetilde{\Gamma}\right) &=& \Omega^{-2} \Bigg\{
 f\left(\Omega\right) \mathbb{R}\left({\Gamma}\right) +
\partial_{\alpha}f\left(\Omega\right) g^{\mu \nu}\Gamma^{\alpha}_{\mu \nu}
 -
\partial_{\nu}f\left(\Omega\right)g^{\mu \nu}\Gamma^{\alpha}_{\mu\alpha} .
\nonumber\\
&+& f \left(\Omega\right) \left(f \left(\Omega\right)-1\right)
\left[\Gamma_{\alpha \lambda}^{\alpha}g^{\mu \nu}\Gamma_{\mu \nu}^{\lambda}-
\Gamma_{\nu \lambda}^{\alpha}g^{\mu \nu}\Gamma_{\mu \alpha}^{\lambda} \right]
\Bigg\} \,.
\end{eqnarray}
It is straightforward to check that the $\Gamma$--dependent terms at the
right-hand side form a true scalar under general coordinate transformations
(See Appendix B for details). This conformal transformation rule for
Ricci scalar dictates what form the
gravitational action (\ref{affine-inv-ac}) in $(g,\Gamma)$ frame takes in
$(\widetilde{g},\widetilde{\Gamma})$
frame. It is clear that the transformed action will involve $\Omega$ as well as
its partial derivatives. Therefore,
contrary to the previous case of conformal-invariant connection, $\Omega$ is a
dynamical field. However,
it does not possess a true kinetic term in the sense of a scalar field theory.
Its derivative interactions
are always accompanied by the connection, $\Gamma^{\lambda}_{\alpha \beta}$.

As another transformation property of the connection, we now turn to analysis of
additive transformation of $\Gamma^{\lambda}_{\alpha \beta}$. We
thus consider the generic transformation rule
\begin{eqnarray}
\label{trans-conn-additive}
\widetilde{\Gamma}^{\lambda}_{\alpha\beta} = \Gamma^{\lambda}_{\alpha\beta}
+
\Delta^{\lambda}_{\alpha\beta}\left(\Omega\right)
\end{eqnarray}
where $\Delta^{\lambda}_{\alpha\beta}\left(\Omega\right)$, being the difference
between $\widetilde{\Gamma}^{\lambda}_{\alpha\beta}$ and
$\Gamma^{\lambda}_{\alpha\beta}$, is a rank (1,2) tensor field. It is a
tensorial connection. This transformation
of the connection is understood to run simultaneously with the transformation of
the metric in (\ref{transmet}). (One notes that $\Delta^{\lambda}_{\alpha\beta}$
here may be interpreted contain a set of vector fields like the conformal factor
itself is a scalar field. See, \cite{yeni-is} for details of such a reduction.) Then, as follows from (\ref{riemann}),
the Riemann tensor transforms as
\begin{eqnarray}
{\widetilde{\mathbb{R}}}^{\alpha}_{\mu \beta
\nu}\left(\widetilde{\Gamma}\right)
= {\mathbb{R}}^{\alpha}_{\mu \beta \nu}\left(\Gamma\right)
+ \nabla_{\beta} \Delta^{\alpha}_{\mu \nu}
- \nabla_{\nu} \Delta^{\alpha}_{\mu \beta}
+ \Delta^{\alpha}_{\lambda \beta} \Delta^{\lambda}_{\mu \nu}
- \Delta^{\alpha}_{\lambda \nu} \Delta^{\lambda}_{\beta \mu}
\label{trans-riemann-additive}
\end{eqnarray}
where the $\Delta$--dependent part at the right-hand side, though seems
so, is not a true curvature tensor; it is not generated by any of
the covariant derivatives induced by $\Gamma^{\lambda}_{\alpha \beta}$
or $\check{\Gamma}^{\lambda}_{\alpha\beta}$. This extra $\Delta$--dependent
piece
is just a rank (1,3) tensor field induced by $\Delta^{\lambda}_{\alpha\beta}$
alone.

In accordance with the transformation of Riemann tensor in
(\ref{trans-riemann-additive}),
the Ricci scalar transforms as
\begin{eqnarray}
\widetilde{g}^{\mu \nu} \widetilde{\mathbb{R}}_{\mu
\nu}\left(\widetilde{\Gamma}\right) = \Omega^{-2} g^{\mu \nu}\left\{
\mathbb{R}_{\mu \nu}\left(\Gamma\right) + \nabla_{\alpha} \Delta_{\mu
\nu}^\alpha
- \nabla_{\nu} \Delta_{\mu \alpha}^\alpha
+ \Delta_{\lambda \alpha}^\alpha \Delta_{\mu \nu}^\lambda
- \Delta_{\lambda \nu}^\alpha \Delta_{\alpha \mu}^\lambda  \right\}\,.
\nonumber\\
\end{eqnarray}
This transformation rule is rather generic for connections which transform
additively \cite{demir}. Nevertheless, it is necessary to determine physically
admissible forms of $\Delta^{\lambda}_{\alpha\beta}$ so that the
conformal factor $\Omega$ assumes appropriate dynamics in regard to absence
of ghosts and emerging of a new conformal invariance in the sense of
(\ref{transmetphix}).

At this stage, right question to ask is this: `How is
$\Delta^{\lambda}_{\alpha\beta}$
related to $\Omega$ ?' To answer this question, one has to check out a series of
possibilities. Being a rank (1,2) tensor field, $\Delta^{\lambda}_{\alpha\beta}$
can
assume a number of forms like $V^{\lambda} g_{\alpha \beta}$ or
$\delta^{\lambda}_{\alpha} V_{\beta}$
or $V^{\lambda} T_{\alpha\beta}$, with $V_{\alpha}$ being a vector field and
$T_{\alpha\beta}$ a
symmetric tensor field. If the transformation of connection
(\ref{trans-conn-additive}) is to
coexist with that of the metric in (\ref{transmet}), then $V_{\alpha}$,
$T_{\alpha\beta}$ or any other
structure must be related to gradients of $\Omega$ so that
$\Delta^{\lambda}_{\alpha\beta}$ vanishes
when $\Omega$ is unity or, more precisely, constant. Therefore, one may identify
$V_{\alpha}$ with $\partial_{\alpha}\Omega$,
and $T_{\alpha\beta}$ with $\nabla_{\alpha}\partial_{\beta}\Omega$ or
$\partial_{\alpha}\Omega\ \partial_{\beta}\Omega$.
Consequently, $\Delta^{\lambda}_{\alpha\beta}$ should be composed of
$\partial^{\lambda}\Omega\ g_{\alpha\beta}$,
$\delta^{\lambda}_{\alpha} \partial_{\beta}\Omega$ or relevant higher
derivatives of $\Omega$ or higher powers of $\partial_{\alpha}\Omega$.
Hence, at the linear level, $\Delta^{\lambda}_{\alpha\beta}$ must be of the form
\begin{eqnarray}
\label{transofcon}
\Delta^{\lambda}_{\alpha\beta} &= c_1 \left( \delta^\lambda_\alpha
\partial_\beta \ln \Omega
+ \delta^\lambda_\beta \partial_\alpha \ln \Omega \right) + c_2 g_{\alpha\beta}
\partial^\lambda \ln \Omega \label{eq:engenel}
\end{eqnarray}
where $c_1$ and $c_2$ are real constants. In here, one notices that a
very similar form of this connection was also found in \cite{shapiro-odintsov,prescription}
in spacetimes with non-vanishing torsion. In \cite{prescription}, prescription in (\ref{transofcon}) is obtained
by requiring invariance of Lorentz connection under conformal transformations. That work also points out
the possibility of conformal invariant gravitational action. In addition to this, by considering a similar prescription for torsion instead of connection, one can construct a conformally-invariant theory. This option is studied in detail in \cite{prescription1}.

That both metric and connection transform according to an assumed prescription (as given in (\ref{transmet}) and (\ref{transofcon}), respectively) may lead one to conclude that metric and connection do actually depend on each other - not independent quantities as required by the metric-affine gravity. Actually, such a dependence does not need to exist. The situation can be clarified by considering, for example, scalar and fermion fields, comparatively. Indeed, they both transform non-trivially under conformal transformations yet they bear no relationship at all. In this sense, one concludes that their behaviors under conformal transformations do not need to impose an inter-dependence between metric and connection.

One readily notices that the tensorial structures involved in (\ref{transofcon}) are the same as the ones appearing in the transformation of the Levi-Civita connection under conformal
transformations. This is seen from direct comparison of (\ref{eq:engenel}) with
(\ref{Delta-for-Levi-Civita}).
The difference is the generality of (\ref{eq:engenel}) in terms of the constants
$c_1$ and $c_2$ since
$c_1=-c_2=1$ in the transformation (\ref{Delta-for-Levi-Civita}) of the
Levi-Civita connection.
Under the transformation (\ref{eq:engenel}), the metric-affine action
(\ref{affine-inv-ac}) in $\left(g, \Gamma\right)$ frame takes the form
\begin{eqnarray}
S_{EH}\left[\widetilde{g},\widetilde{\Gamma},\overline{\phi}\right]
&=& \int d^D x\ \sqrt{-g} \Bigg\{
\frac{1}{2} \Omega^{D-2} M_\star^{D-2} g^{\mu \nu} \mathbb{R}_{\mu
\nu}\left(\Gamma\right) \nonumber\\
&\quad& + \frac{1}{2} (D-1)(D-2) {\kappa}_D  \Omega^{D-4}
M_\star^{D-2} g^{\mu \nu} \partial_\mu \Omega \partial_\nu \Omega \nonumber \\
&\quad& - \Lambda_{\star} \Omega^D + \widetilde{\mathcal{L}} \Bigg\}\nonumber\\
\label{final-action-additive}
&=& \int d^D x \sqrt{-g} \Bigg\{ \frac{1}{2} \mbox{Sign} \left(\kappa_D\right)
g^{\mu \nu} \partial_{\mu} \overline{\phi} \partial_\nu \overline{\phi}
\nonumber\\
&\quad& + \frac{1}{2} \zeta_D^{\prime} \overline{\phi}^2 g^{\mu\nu}
\mathbb{R}_{\mu\nu}\left(\Gamma\right) - \lambda_D \left(
\zeta_D^{\prime} \overline{\phi}^2 \right)^{\frac{D}{D-2}} \Bigg\}
\end{eqnarray}
where use has been made of the abbreviations
\begin{eqnarray}
\kappa_D &=& \frac{\left(c_1 + c_2 \right)^2  +(D-2) c_1 c_2 + (D-2)
(c_1 - c_2)}{D-2} \\
\zeta_D^{\prime} &=& \frac{\zeta_D}{\left|\kappa_D\right|}
\end{eqnarray}
along with the new canonical scalar field
\begin{eqnarray}
\overline{\phi}=\frac{1}{\sqrt{\zeta_D^{\prime}}} \left(M_{\star}
\Omega\right)^{\frac{D-2}{2}}\,.
\end{eqnarray}
The action (\ref{final-action-additive}) is to be contrasted with the
transformed action (\ref{eq:GR-action})
in metrical gravity. The differences between the two are spectacular, and it
could prove useful to discuss them
here in detail:
\begin{itemize}
\item One first notes that, the action (\ref{final-action-additive}) is
invariant under the emergent
conformal transformations
\begin{eqnarray}
\label{transmetphixx}
&&{g}_{\alpha \beta} \longrightarrow \psi^{2}\ {g}_{\alpha \beta}\nonumber\\
&&\Gamma^{\lambda}_{\alpha\beta} \longrightarrow
\Gamma^{\lambda}_{\alpha\beta} + \Delta^{\lambda}_{\alpha
\beta}(\psi)\nonumber\\
&&\overline{\phi} \longrightarrow \psi^{-\frac{(D-2)}{2}}\ \overline{\phi}
\end{eqnarray}
similar to what we have found in (\ref{transmetphi}) for the metrical GR. This
invariance
guarantees that all the fixed scales in (\ref{affine-inv-ac}) are appropriately
dressed
by the conformal factor $\Omega$ \cite{bekenstein}.

\item The conformal coupling $\zeta_D$ in (\ref{eq:GR-action}) of the pure
metric gravity changes
to $\zeta_D/\left|\kappa_D\right|$ in the metric-affine action under concern.
The presence of
$\kappa_D$ reflects the generality of the transformation of the connection, as
noted in (\ref{eq:engenel}).
This is a highly important result since it generalizes the very concept of
`conformal coupling' between scalar
fields and curvature scalar by changing $\zeta_D$ to $\zeta_D^{\prime}$. This
modification can have observable
consequences in cosmological \cite{bauer1,cosmo1,cosmo2} as well as collider
observables \cite{giudice,demir2} of the GR with affinity.

\item In complete contrast to (\ref{eq:GR-action}), the scalar field
$\overline{\phi}$ in (\ref{final-action-additive})
obtains an indefinite kinetic term. The sign of the kinetic term is determined
by the sign of $\kappa_D$. One here notes two
physically distinct cases:
\begin{enumerate}
\item If $\kappa_D > 0$ then $\overline{\phi}$ is a scalar ghost as in the
metrical GR. In (\ref{eq:GR-action}) $\kappa_D =1$
(since $c_1=1$ and $c_2=-1$ for the change of Levi-Civita connection
(\ref{Delta-for-Levi-Civita}) under
conformal transformations), and $\overline{\phi}$ is necessarily a ghost if
gravity is to stay as an attractive force.

\item If, however, $\kappa_D < 0$ then $\overline{\phi}$ becomes a true scalar
field with no problems like ghosty behavior.
One notices from (\ref{final-action-additive}) that this very regime is realized
with no modification in the attractive nature of the gravitational
force. Gravity is attractive and $\overline{\phi}$ is a non-ghost, true scalar
field. This result follows form the generality
of the transformation of $\Gamma^{\lambda}_{\alpha \beta}$ in
(\ref{eq:engenel}) compared to that of the Levi-Civita connection. The
real constants $c_1$ and $c_2$ gives enough freedom to make $\kappa_D$
negative for having a canonical scalar field theory, and this happens for
\begin{eqnarray}
 c_2 &>& -1 + \frac12 D(1-c_1) \nonumber - \frac12  \sqrt{ (D^2-4)
c_1^2 - 2(D^2-4) c_1 + (D-2)^2 }
\end{eqnarray}
and
\begin{eqnarray}
 c_2 &<& -1 + \frac12 D(1-c_1) \nonumber + \frac12  \sqrt{ (D^2-4)
c_1^2 - 2(D^2-4) c_1 + (D-2)^2 }
\end{eqnarray}
where $c_1$ is restricted to lie outside the interval $\left[ \frac{D+2 - 2
\sqrt{D+2}}{(D+2)} \: ,
\frac{D+2 + 2
\sqrt{D+2}}{(D+2)} \right]$. One can see that for any
dimension
$D\geq 4$ there exist wide ranges of values of $c_1$ for which $c_2$ takes on
admissible negative or positive real values. In particular, if we consider one of the
most likly cases in which the constants are equal but have
opposite signs, we find $\kappa_D < 0$ for $c_1 = -c_2 \not\in \left( 0, 2 \right]$
in $D > 2$ dimensions. Similar considerations pertaining
to the metric-scalar-torsion system can be found in \cite{neto}.

\item The fact that the metric-affine gravity offers a true scalar field
$\overline{\phi}$
elevates the arguments on the cosmological constant problem in \cite{polyakov}
to a more physical
status since one then does not need to multiply the scalar field by the
imaginary unit to
make sense of the resulting scalar field theory. For $\kappa_D < 0$ and
$\tilde{g}_{\mu
\nu} = \eta_{\mu\nu}$, the affine-gravitational action
(\ref{final-action-additive}) can
realize infrared fixed point for $\overline{\phi}$ with no artificial changes in
the sign of its
kinetic term.

\end{enumerate}

\item The geometrical part of $\mathcal{L} \left( g, \mathcal{D},
\Psi\right)$, which only consist of the metric and
$\mathcal{D}_{\alpha \beta}^\lambda = \Gamma_{\alpha \beta}^\lambda -
\check{\Gamma}_{\alpha \beta}^\lambda $, will also transform under
conformal transformation (\ref{transmet}) with additively conformal-variant
connection (\ref{trans-conn-additive}). Under the conformal transformations
(\ref{transmetphixx}), $\cal{D}$
changes as
\begin{eqnarray}
\widetilde{\cal{D}}^{\lambda}_{\alpha\beta} &=&
{\cal{D}}^{\lambda}_{\alpha\beta} + (c_2 + 1) g_{\alpha\beta}
\partial^\lambda \ln \psi
+(c_1 - 1) \left( \delta^\lambda_\alpha \partial_\beta \ln \psi
+ \delta^\lambda_\beta \partial_\alpha \ln \psi \right) \nonumber
\end{eqnarray}
as expected from transformation properties of $\Gamma^{\lambda}_{\alpha\beta}$
and $\check{\Gamma}^{\lambda}_{\alpha\beta}$. This gives geometrodynamical terms
and
couplings of $\mathcal{D}$ with the emergent scalar field ${\psi}$.

\item An important problem concerns the gravitational kinetic term. In metric
formulation, conformally-invariant kinetic term is provided by the
Weyl tensor \cite{demir,zee}. In the present case, however, such a
conformal-invariant kinetic term might be difficult to induce. Actually,
what is necessary is to construct a gravitational kinetic term which
is invariant under the conformal transformations in (\ref{transmet}) and
(\ref{trans-conn-additive}) (with the specific form in (\ref{transofcon})).
With $c_1$ and $c_2$ differing from the Levi-Civita connection, construction
of the kinetic term may not be straightforward.

\end{itemize}

The analysis above ensures that additively transforming connections, such as the
one (\ref{eq:engenel}), gives
rise to a physically sensible mechanism where gravitational sector as well as
the emergent scalar field
from conformal transformation are both physical. Removal of the ghosty degree of
freedom in metrical GR is
a highly important aspect of the metric-affine gravity. Essentially, freeing
connection from metric enables
one to reach a physically consistent picture in regard to conformal frame
changes in the gravitational action.

\section{Discussions and Conclusion}
\label{sec:conc}

In this work we have analyzed conformal transformations in metric-affine
gravity (GR with affinity). The analysis
is a comparative one between the GR with affinity and metricity. The main result
of the analysis is that metric-affine
gravity admits, under general additive transformations of the connection,
conformally-related frames in which
both gravitational and scalar sectors behave physically. The transformed frame
consists of no ghost field, and
exhibits emergent conformal invariance (sometimes called Weyl-St{\"u}ckelberg
invariance). The results can have
far-reaching consequences for collider experiments \cite{giudice,demir2},
cosmological evolution \cite{bauer1,application,higher-deriv} as well as
the electroweak breaking \cite{demir}.

We have also analyzed equations of motion under general circumstances allowed by
general covariance, and concluded
that general Lagrangians allow for  generalized conformal transformations of the
connection without spoiling
the essence of the theory in the transformed frame.

The affine gravitational action (\ref{affine-inv-ac}) can give rise to novel
effects not found in the minimal
version (the Einstein-Hilbert action). The conformally-reached frame can have
various modifications in gravitational,
matter as well as conformal factor ({\it i.e.} the $\Omega$ related to
$\overline{\phi}$) dynamics. The fact that
the metric-affine gravity can accommodate correct gravitational dynamics plus
non-ghost scalar degree of freedom under
conformal transformations is an  important aspect. This feature can have
important implications in cosmological and other settings since transformation
of system to a conformal frame now involves no ghosty degree of freedom . Indeed,
the appearance of  ghost fields, as mentioned in the text, is the major problem of conformal general relativity
\cite{nobili}. Therefore, the ghost-free dynamics established in the present work can have significant
applications in conformal field theory, cosmology and gravitation.

\section{Acknowledgements}
The authors gratefully thank Bar{\i}{\c s} Ate{\c s} for his mathematical
contributions at the early stages of this work. Authors also thank to
very conscientious referees for their comments and suggestions.

\section{Appendices}

\subsection{A. Equations of motion}
\label{sec:eom}

We have found that metric-affine gravity provides a means of generating
non-ghost scalar field $\overline{\phi}$
by executing a more general transformation property as indicated in
(\ref{eq:engenel}). However, we know that
equations of motion relate $\Gamma^{\lambda}_{\alpha \beta}$ to Levi-Civita
connection, and it is questionable if one
can indeed realize such generalized transformation properties. For a detailed
analysis of the problem, we will
proceed systematically by examining different forms of geometrodynamical action
densities.

\begin{itemize}
\item First of all, one notes that the affine gravitational action
(\ref{affine-inv-ac}) becomes a highly
conservative one for $\mathcal{L} = 0$. In this case, variation of action with
respect to the connection
$\Gamma^{\lambda}_{\alpha \beta}$ gives
\begin{eqnarray}
\label{equival}
\nabla^{\Gamma}_{\lambda} \left( \sqrt{-g} g^{\alpha \beta} \right) = 0
\end{eqnarray}
where the covariant derivative of the tensor densitiy is defined as
\begin{eqnarray}
 \nabla^{\Gamma}_{\lambda} \sqrt{-g} = \partial_\lambda \sqrt{-g} -
\Gamma_{\alpha \lambda}^\alpha \sqrt{-g}
\end{eqnarray}
Then the equation (\ref{equival}) is solved uniquely for
\begin{eqnarray}
\Gamma^{\lambda}_{\alpha\beta} = \check{\Gamma}^{\lambda}_{\alpha\beta}.
\end{eqnarray}
Therefore, the action (\ref{affine-inv-ac}) is equivalent to the action for
metrical gravity
in (\ref{eq:grav-ac}). The main advantage of metric-affine gravity (actually the
Palatini formalism
itself) is that one arrives at the equations of GR with no need to extrinsic
curvature
(which is needed in metrical gravity). In sum, with $\mathcal{L} = 0$,
(\ref{affine-inv-ac}) gives
an equivalent description of (\ref{eq:grav-ac}). We will elaborate more on this
point below.

\item  There can, however, be various sources of departure from the
action (\ref{affine-inv-ac}). These sources of departure are contained in
$\mathcal{L}$. Let us
first examine $\mathcal{L}_{\mbox{geo}} \left( g, \mathcal{D}
\right)$ which involves metric and the tensorial connection
$\mathcal{D}^{\lambda}_{\alpha\beta}$.
The tensorial connection $\mathcal{D}^{\lambda}_{\alpha\beta}$ gives rise to
novel geometrodynamical
structures not necessarily governed by the curvature tensor
${\mathbb{R}}^{\alpha}_{\mu \beta \nu}\left(\Gamma\right)$
and its contractions and higher powers (though such sources of
$\mathcal{D}^{\lambda}_{\alpha \beta}$ are to be also included
in $ \mathcal{L}_{\mbox{geo}} \left( g, \mathcal{D}
\right)$). Indeed, the action can be added various new terms involving
appropriate powers of
${\cal{D}}^{\lambda}_{\alpha\beta}$ as long as general covariance is respected.
One notices that
only even powers of ${\cal{D}}^{\lambda}_{\alpha \beta}$ can arise in the action
\cite{recent}. Needless to say, presence of additional terms
involving ${\cal{D}}^{\lambda}_{\alpha \beta}$ changes the equation of motion
for
$\Gamma^{\lambda}_{\alpha\beta}$. In particular, its dynamical
equivalence to Levi-Civita connection, in the sense of (\ref{equival}), gets
lost.

For explicating these points we go back to (\ref{affine-inv-ac}) and switch on
$\mathcal{L}_{\mbox{geo}} \left( g, \mathcal{D} \right)$  after which the
${\cal{D}}^{\lambda}_{\alpha \beta}$ dependence of the action takes the form
\begin{eqnarray}
S_{EH}\left[g, \mathcal{D} \right] &=& \int d^{D} x\ \sqrt{-g} \Bigg\{
\frac{1}{2}
M_{\star}^{D-2} g^{\mu \nu} \underbrace{ \left[ R_{\mu
\nu}\left(\check{\Gamma}\right) +
\mathcal{R}_{\mu \nu}\left(\mathcal{D}\right) \right]}_{\mathbb{R}_{\mu\nu}
\left(\Gamma\right) } \nonumber\\
 &-& \Lambda_{\star} + \mathcal{L}_{\mbox{geo}} \left( g, \mathcal{D} \right)
\Bigg\} \label{affine-inv-ac2}
\end{eqnarray}
where we discarded $\mathcal{L}\left( g, \mathcal{D},
\psi \right)$ momentarily, to analyze the effects of geometrical
part of $\mathcal{L}$ in isolation. Actually, as we have mentioned before in (\ref{express}),
$\mathcal{L}_{\mbox{geo}}\left( \mathcal{D} \right)$ can always be expressed in terms of
torsion (which vanishes in our case), non-metricity, and curvature tensors. A more general discussion of the roles of non-metricity and torsion could be found in \cite{Sotiriou:2009}. We here prefer to use generic function $\mathcal{L}_{\mbox{geo}} \left( \mathcal{D} \right)$
instead of expressing it in terms of those tensor structures in (\ref{express}). From
(\ref{trans-riemann-additive}) it follows
that
\begin{eqnarray}
\mathcal{R}_{\mu \nu} \left( \mathcal{D} \right) &=& \nabla_{\alpha}
\mathcal{D}_{\mu \nu}^\alpha - \nabla_{\nu} \mathcal{D}_{\mu \alpha}^\alpha
+ \mathcal{D}_{\lambda \alpha}^\alpha \mathcal{D}_{\mu
\nu}^\lambda
- \mathcal{D}_{\lambda \nu}^\alpha \mathcal{D}_{\alpha \mu}^\lambda
\end{eqnarray}
in the action (\ref{affine-inv-ac2}). Variation of the action with respect to
$\mathcal{D}^{\lambda}_{\alpha\beta}(z)$
gives the equations of motion
\begin{eqnarray}
\delta_\lambda^\beta g^{\mu \nu} (z)
\mathcal{D}_{\mu\nu}^\alpha + g^{\alpha \beta} (z) \mathcal{D}_{\lambda \nu}^\nu
(z) - g^{\mu \beta} (z) \mathcal{D}_{\lambda \mu}^\alpha (z) - g^{\beta \nu}
(z) \mathcal{D}_{\lambda \nu}^\alpha (z) + \mathcal{G}_\lambda^{\alpha \beta}
\left( g, \mathcal{D} \right) = 0 \nonumber\\
\label{eq:eqofmot}
\end{eqnarray}
where $\mathcal{G}_\lambda^{\alpha \beta} \left( g, \mathcal{D} \right)$ stands
for the variation of the geometrical part $\mathcal{L}_{\mbox{geo}}\left( g,
\mathcal{D} \right)$ with respect to $\mathcal{D}^{\lambda}_{\alpha\beta}(z)$.

\begin{itemize}
\item
From (\ref{eq:eqofmot}) one immediately observes that, for
$\mathcal{L}_{\mbox{geo}} \left( g,
\mathcal{D} \right) = 0$ (in addition to assumed vanishing of the matter
contribution), the
tensorial connection identically vanishes, $\mathcal{D}^{\lambda}_{\alpha\beta}
= 0$. This
implies that the general connection $\Gamma^{\lambda}_{\alpha \beta}$ equals
the Levi-Civita
connection $\check{\Gamma}^{\lambda}_{\alpha\beta}$. In such a case, of course,
$\Gamma^{\lambda}_{\alpha \beta}$
is expected to exhibit the same transformation properties as
$\check{\Gamma}^{\lambda}_{\alpha\beta}$. Consequently,
the general conformal transformation property (\ref{eq:engenel}) as well as the
conclusions drawn from
it will not hold for minimal Lagrangians, like (\ref{affine-inv-ac}) with
$\mathcal{L}=0$. In this
sense, analysis of the previous section, though designed to show how varying
conformal transformation
properties of $\Gamma^{\lambda}_{\alpha \beta}$ modify the ghosty nature of
$\overline{\phi}$, is physically
sensible yet incomplete for it does not take into  account the effects of
non-vanishing $\mathcal{L}$ effects.

\item We have just concluded that we need non-vanishing $\mathcal{L}$ for
maintaining the independence of $\Gamma^{\lambda}_{\alpha\beta}$
from $\check{\Gamma}^{\lambda}_{\alpha\beta}$. Now it proves useful to check
some
reasonable forms of $\mathcal{L}_{\mbox{geo}} \left( g,
\mathcal{D} \right)$ in light of the equations of motion (\ref{eq:eqofmot}).
Leaving aside the single-derivative
terms as well as quadratic ones whose special forms are already contained in the
curvature tensor, the lowest-order
terms which can contribute to geometrical part take the form
\begin{eqnarray}
\label{formx}
\mathcal{L}_{\mbox{geo}} \left( g, \mathcal{D} \right)
&=& A_{\lambda \rho \zeta \epsilon}^{\alpha \beta \mu \nu \chi \xi \eta \kappa}
\left( g
\right)
\mathcal{D}_{\alpha \beta}^\lambda \mathcal{D}_{\mu\nu}^\rho
\mathcal{D}_{\chi \xi}^\zeta \mathcal{D}_{\eta \kappa}^\epsilon
\left( \mathcal{D}
\right)
\nonumber \\ &+&
B_{\lambda \zeta}^{\alpha \beta \mu \nu \rho \theta} \left( g \right)
\nabla_\mu \mathcal{D}_{\alpha \beta}^\lambda
\nabla_\nu \mathcal{D}_{\rho \theta}^\zeta
+ \cdots
\end{eqnarray}
where $A$ and $B$ are tensorial structures composed of the metric tensor. They
are supposed to contain all possible combinatorics of the indices. It is clear
that, after computing  $\mathcal{G}_\lambda^{\alpha
\beta} \left( g, \mathcal{D} \right)$ from this combination, the equations
of motion (\ref{eq:eqofmot}) will yield non-vanishing
${\cal{D}}^{\lambda}_{\alpha\beta}$
even without including its derivatives. Indeed, having (\ref{formx}) at hand,
the
equations of motion (\ref{eq:eqofmot}) take the form
\begin{eqnarray}
\label{eq:eomfordiff}
 && \mathcal{D}_{\rho \theta}^\sigma \left[
g^{\rho \theta} \delta_\lambda^\beta \delta_\sigma^\alpha
+ g^{\alpha \beta} \delta_\sigma^\theta \delta_\lambda^\rho
- g^{\theta \beta} \delta_\sigma^\alpha \delta_\lambda^\rho
- g^{\beta \theta} \delta_\sigma^\alpha \delta_\lambda^\rho
\right. \nonumber \\
&&+ \left. \mathcal{D}_{\chi \xi}^\zeta \mathcal{D}_{\eta \kappa}^\epsilon
\left( A_{\lambda \sigma \zeta \epsilon}^{\alpha \beta \rho \theta \chi \xi
\eta \kappa}
+ A_{\lambda \sigma \zeta \epsilon}^{\rho \theta \alpha \beta \chi \xi \eta
\kappa}\right) \right. \nonumber \\ &&+ \left. \mathcal{D}_{\mu \nu}^\zeta
\mathcal{D}_{\chi \xi}^\epsilon
\left( A_{\sigma \zeta \lambda \epsilon}^{\rho \theta \mu \nu \alpha \beta
\chi \xi}
+  A_{\sigma \zeta \epsilon \lambda}^{\rho \theta \mu \nu \chi \xi \alpha
\beta} \right) \right] \nonumber \\
&-& \nabla_\rho \nabla_\theta \mathcal{D}_{\mu \nu}^\sigma
\left(  B_{\lambda \sigma}^{\rho \theta \alpha \beta \mu \nu} +
B_{\lambda \sigma}^{\rho \theta \mu \nu \alpha \beta} \right) = 0 \,.
\end{eqnarray}
These equations automatically suggest that $\mathcal{D}^{\lambda}_{\alpha\beta}
\neq 0$ (or
$\Gamma^{\lambda}_{\alpha\beta}\neq \check{\Gamma}^{\lambda}_{\alpha\beta}$)
even if
$\mathcal{L}_{\mbox{geo}} \left( g,
\mathcal{D} \right)$ does not include its derivatives (the coefficients $B$
vanish). If derivative terms
vanish, then $\mathcal{D}^{\lambda}_{\alpha\beta}$ is obtained in terms of the
metric
tensor with, however, a general structure which should resemble
(\ref{eq:engenel}) in
any case. The details of the structure depend on how the coefficients
$A_{\lambda \rho \zeta \epsilon}^{\alpha \beta \mu \nu \chi \xi \eta \kappa}$
are organized in terms of the metric tensor.

On the other hand, if the derivative terms are included then $
\mathcal{D}^{\lambda}_{\alpha\beta}$
becomes a dynamical field. In this case, again, one obtains a non-trivial
$\Gamma^{\lambda}_{\alpha\beta}$
not equaling $\check{\Gamma}^{\lambda}_{\alpha\beta}$.

A simple illustrative example for the aforementioned Lagrangian would be the
geometrical quantity
\begin{eqnarray}
 \mathcal{L}_{\mbox{geo}} \left( g, \mathcal{D} \right) &=& a
\mathbb{C}_\alpha^{\mu\beta\nu} \mathbb{C}^\alpha_{\mu\beta\nu}
\end{eqnarray}
where $a$ is a suitable constant of mass dimension $D-4$, and
$\mathbb{C}^\alpha_{\mu\beta\nu} \left( \Gamma \right)$ is the $D$-dimensional Weyl curvature tensor \cite{cosmo1}.
It is nothing but the traceless part of the Riemann curvature tensor
$\mathbb{R}^{\mu}_{\alpha\nu\beta} \left( \Gamma \right)$, and has the same index symmetries.
The resulting field equations will be of the form (\ref{eq:eomfordiff}) which tells
us that the tensorial connection $\mathcal{D}^\lambda_{\mu\nu}$ is an independent
dynamical field. (Discussions of these points can be found in \cite{Vitagliano:2010sr} for
the case with non-vanishing torsion.)

From this analysis we conclude that, the analysis of the previous section, which
has clearly shown how
$\overline{\phi}$ becomes a non-ghost scalar for a general
$\Gamma^{\lambda}_{\alpha\beta}$ transforming
as in (\ref{eq:engenel}), in general, the connection
$\Gamma^{\lambda}_{\alpha\beta}$ does not reduce to
$\check{\Gamma}^{\lambda}_{\alpha\beta}$, and a conformal transformation
property as in
(\ref{eq:engenel}) can
result in a multitude of ways.

\item Another source of departure from (\ref{affine-inv-ac}) is the matter
Lagrangian $\mathcal{L}\left( g, \mathcal{D}, \psi \right)$. By switching on this
Lagrangian one can still find additional
structures which cause $\Gamma^{\lambda}_{\alpha\beta}$ to be independent of
$\check{\Gamma}^{\lambda}_{\alpha\beta}$.
Then the main difference from the previous analysis will be the dependence of
the $\Gamma^{\lambda}_{\alpha\beta}$
on the matter fields themselves -- a situation not discussed before. The
question of how $\mathcal{L}
\left( g, \mathcal{D}, \psi \right)$ involves
$\Gamma^{\lambda}_{\alpha\beta}$ is easy to answer
given that, rather generically, connection-dependent terms arise in scalar and spinor field theories already at
the renormalizable level \cite{deser1}. In such cases it could be difficult to arrange general conformal transformations
of the form (\ref{eq:engenel}) yet one should keep such matter sector sources in mind in analyzing the conformal
transformation properties in non-Riemannian geometries.

\end{itemize}
\end{itemize}

\subsection{B. Ricci Scalar Under Multiplicatively Transforming Connection}
In this appendix, we sketch the calculations showing that the multiplicatively
transformed Ricci ``scalar'' $\tilde{g}^{\mu\nu}
\tilde{\mathbb{R}} (\Gamma)_{\mu\nu}$ is a true scalar under general
coordinate transformations. Indeed, by direct calculation, one finds
step by step the following relations:
\begin{eqnarray*}
&&\Omega^{-2} \partial_\alpha f\left( \Omega
\right)\left(g^{\mu\nu}\Gamma_{\mu\nu}^\alpha -
g^{\mu\alpha} \Gamma_{\mu\lambda}^\lambda \right) \\
&&+ \Omega^{-2} \left[ f^2
\left( \Omega \right)  -f \left( \Omega \right) \right] \left( \Gamma_{\alpha
\lambda}^\alpha g^{\mu\nu} \Gamma_{\mu\nu}^\lambda - \Gamma_{\nu\lambda}^\alpha
g^{\mu\nu} \Gamma_{\mu\alpha}^\lambda \right)\\
&\Rightarrow& \Omega^{-2} \partial_{\alpha} f(\Omega) \Bigg[
\frac{\partial x^{\mu}}{\partial x^{\mu^\prime}} \frac{\partial
x^{\nu}}{\partial x^{\nu^\prime}} g^{\mu^\prime \nu^\prime} \left(
\frac{\partial x^{\alpha}}{\partial x^{\alpha^\prime}} \frac{\partial
x^{\mu^\prime}}{\partial x^{\mu}} \frac{\partial x^{\nu^\prime}}{\partial
x^{\nu}} \Gamma_{\mu^\prime \nu^\prime}^{\alpha^\prime} + \frac{\partial
x^{\alpha}}{\partial
x^{\alpha^\prime}}\frac{\partial^{2}x^{\alpha^\prime}}{\partial x^{\mu }
\partial x^{\nu}}\right) \\
&&- \frac{\partial x^{\mu}}{\partial x^{\mu^\prime}} \frac{\partial
x^{\alpha}}{\partial x^{\alpha^\prime}} g^{\mu^\prime
\alpha^\prime} \left( \frac{\partial x^{\lambda}}{\partial
x^{\lambda^\prime}} \frac{\partial x^{\lambda^\prime}}{\partial
x^{\lambda}} \frac{\partial x^{\mu^\prime}}{\partial x^{\mu}}
\Gamma_{\lambda^\prime \mu^\prime}^{\lambda^\prime} + \frac{\partial
x^{\lambda}}{\partial x^{\lambda^\prime}} \frac{\partial^{2}
x^{\lambda^\prime}}{\partial x^{\lambda} \partial x^{\mu}} \right) \Bigg] \\
&&+ \Omega^{-2}(f^{2}-f) \Bigg[ \left(\frac{\partial x^{\alpha}}{\partial
x^{\alpha^\prime}}\frac{\partial x^{\alpha^\prime}}{\partial
x^{\alpha}}\frac{\partial x^{\lambda^\prime}}{\partial
x^{\lambda}}\Gamma_{\alpha^\prime \lambda^\prime}^{\alpha^\prime}+\frac{\partial
x^{\alpha}}{\partial x^{\alpha^\prime}}\frac{\partial^2
x^{\alpha^\prime}}{\partial x^{\alpha} \partial x^{\lambda}}\right) \\
&& \cdot \frac{\partial x^{\mu}}{\partial x^{\mu^\prime}}\frac{\partial
x^{\nu}}{\partial x^{\nu^\prime}} g^{\mu^\prime \nu^\prime} \left(
\frac{\partial x^\lambda }{\partial x^{\lambda^\prime}}
\frac{\partial x^{\mu^\prime}}{\partial x^\mu }
\frac{\partial x^{\nu^\prime}}{\partial x^\nu}
\Gamma_{\mu^\prime \nu^\prime}^{\lambda^\prime}
+ \frac{\partial x^\lambda}{\partial x^{\lambda^\prime}}
\frac{\partial^2 x^{\lambda^\prime} }{ \partial x^\mu \partial x^\nu }
\right) \nonumber\\
&&- \left(\frac{\partial x^{\alpha}}{\partial x^{\alpha^\prime}}\frac{\partial
x^{\nu^\prime}}{\partial x^{\nu}}\frac{\partial x^{\lambda^\prime}}{\partial
x^{\lambda}}\Gamma_{\nu^\prime \lambda^\prime}^{\alpha^\prime}+\frac{\partial
x^{\alpha}}{\partial x^{\alpha^\prime}}\frac{\partial^{2}
x^{\alpha^\prime}}{\partial x^{\nu}\partial x^{\lambda}}\right) \\
&& \cdot \frac{\partial
x^{\mu}}{\partial x^{\mu^\prime}}\frac{\partial x^{\nu}}{\partial
x^{\nu^\prime}}g^{\mu^\prime \nu^\prime}
\left(
\frac{\partial x^{\lambda}}{\partial x^{\lambda^\prime}}
\frac{\partial x^{\mu^\prime}}{\partial x^{\mu}}
\frac{\partial x^{\alpha^\prime}}{\partial x^{\alpha}}
\Gamma_{\mu^\prime \alpha^\prime}^{\lambda^\prime}
+\frac{\partial x^{\lambda}}{\partial x^{\lambda\prime}}
\frac{\partial^{2} x^{\lambda^\prime}}{\partial x^\mu \partial x^\alpha}
\right) \Bigg] \\
&=& \Omega^{-2}\partial_{\alpha}f(\Omega) \Bigg[
\frac{\partial x^{\alpha}}{\partial x^{\alpha^\prime}}
g^{\mu^\prime \nu^\prime}
\Gamma_{\mu^\prime \nu^\prime}^{\alpha^\prime}
+\frac{\partial x^{\alpha}}{\alpha^\prime}
 \frac{\partial x^{\mu}}{\partial x^{\mu^\prime}}
\frac{\partial x^{\nu}}{\partial x^{\nu^\prime}}
\frac{\partial}{\partial x^{\nu}}
\left(
\frac{\partial x^{\alpha^\prime}}{\partial x^{\mu}}
\right)
g^{\mu^\prime \alpha^\prime}\\
&&- \frac{\partial x^{\alpha}}{\partial x^{\alpha^\prime}}
g^{\mu^\prime \alpha^\prime}
\Gamma_{\lambda^\prime \mu^\prime}^{\lambda^\prime}
-\frac{\partial x^{\alpha}}{\partial x^{\alpha^\prime}}
\frac{\partial x^{\lambda}}{\partial x^{\lambda^\prime}}
g^{\mu^\prime \nu^\prime}
\frac{\partial x^{\mu}}{\partial x^{\mu^\prime}}
\frac{\partial^{2} x^{\lambda^\prime}}{\partial x^{\lambda} \partial x^{\mu}}
\Bigg] \\
&&+ \Omega^{-2} (f^2 - f)
\left[
\left(
\frac{\partial x^{\lambda^\prime}}{\partial x^{\lambda}}
\Gamma_{\alpha^\prime \lambda\prime}
\right)
\left( g^{\mu^\prime \nu^\prime}
\frac{\partial x^{\lambda}}{\partial x^{\lambda^\prime}}
\Gamma_{\mu^\prime \nu^\prime}^{\lambda^\prime}
+  \frac{\partial x^{\lambda}}{\partial x^{\lambda^\prime}}
\frac{\partial x^\mu}{\partial x^{\mu^\prime} }
g^{\mu^\prime \nu^\prime}
\frac{\partial x^{\nu}}{\partial x^{\nu^\prime}}
\frac{\partial^2 x^{\lambda^\prime}}{\partial x^\mu \partial x^{\nu}}
\right)\right]\\
&&- \Omega^{-2}(f^2-f) \Bigg[
\left(
\frac{\partial x^{\alpha}}{\partial x^{\alpha^\prime}}
\frac{\partial x^{\lambda^\prime}}{\partial x^{\lambda}}
\Gamma_{\nu^\prime \lambda^\prime}^{\alpha^\prime}
+\frac{\partial x^{\alpha}}{\partial x^{\alpha^\prime}}
\frac{\partial x^{\nu}}{\partial x^{\nu^\prime}}
\frac{\partial^2 x^{\alpha^\prime}}{\partial x^{\nu}\partial x^{\lambda}}
\right)\\
&& \cdot \left(
g^{\mu^\prime \nu^\prime}
\frac{\partial x^{\lambda}}{\partial x^{\lambda^\prime}}
\frac{\partial x^{\alpha^\prime}}{\partial x^{\alpha}}
\Gamma_{\mu^\prime \alpha^\prime}^{\lambda^\prime}
+ g^{\mu^\prime \nu^\prime}
\frac{\partial x^{\mu}}{\partial x^{\mu^\prime}}
\frac{\partial x^{\lambda}}{\partial x^{\lambda^\prime}}
\frac{\partial^2 x^{\lambda^\prime}}{\partial x^{\mu} \partial x^{\alpha}}
\right)
\Bigg]\\
&=& \Omega^{-2}
\frac{\partial f}{\partial x^{\alpha^\prime}}
\frac{\partial x^{\alpha^\prime}}{\partial x^{\alpha}}
\frac{\partial x^{\alpha}}{\partial x^{\alpha^\prime}}
g^{\mu^\prime \nu^\prime}
\Gamma_{\mu^\prime \nu^\prime}^{\alpha^\prime}
- \Omega^{-2}
\frac{\partial f}{\partial x^{\alpha^\prime}}
\frac{\partial x^{\alpha^\prime}}{\partial x^{\alpha}}
\frac{\partial x^{\alpha}}{\partial x^{\alpha^\prime}}
g^{\mu^\prime \alpha^\prime}
\Gamma_{\mu^\prime \lambda^\prime}^{\lambda^\prime} \\
&&+ \Omega^{-2} (f^2-f) \left[ \Gamma_{\alpha^\prime
\lambda^\prime}^{\alpha^\prime} g^{\mu^\prime \nu^\prime} \Gamma_{\mu^\prime
\nu^\prime}^{\lambda^\prime} - \Gamma_{\nu^\prime
\lambda^\prime}^{\alpha^\prime} g^{\mu^\prime \nu^\prime} \Gamma_{\mu^\prime
\alpha^\prime}^{\lambda^\prime} \right]\\
&=& \Omega^{-2} \partial_{\alpha^\prime} f (\Omega) g^{\mu^\prime \nu^\prime}
\Gamma_{\mu^\prime \nu^{\prime}}^{\alpha^\prime}-
\Omega^{-2} \partial_{\nu^\prime} f (\Omega) g^{\mu^\prime
\nu^\prime} \Gamma_{\alpha^\prime \mu^\prime}^{\alpha^\prime}\\
&&+ \Omega^{-2} (f^2-f) \left[ \Gamma_{\alpha^\prime
\lambda^\prime}^{\alpha^\prime} g^{\mu^\prime \nu^\prime} \Gamma_{\mu^\prime
\nu^\prime}^{\lambda^\prime} - \Gamma_{\nu^\prime
\lambda^\prime}^{\alpha\prime}g^{\mu^\prime \nu^\prime} \Gamma_{\mu^\prime
\alpha^\prime}^{\lambda^\prime} \right]
\end{eqnarray*}
which is precisely what is to be shown.

\end{document}